\RequirePackage{lineno}
\documentclass[sort&compress,twocolumn,10pt,5p]{elsarticle}

\usepackage{atlasphysics}
\pdfoutput=1
\DeclareGraphicsExtensions{.pdf}

\usepackage{subfigure}
\usepackage{mathrsfs}
\usepackage{graphicx}
\usepackage{multirow}
\usepackage{rotating}
\usepackage{xspace}
\usepackage{slashed}
\usepackage{amsmath}
\usepackage{hyperref}

\newcommand{\alpgen}{{\tt ALPGEN}}
\newcommand{\powheg}{{\tt POWHEG}}

\newcommand{\herwig}{{\tt HERWIG}}
\newcommand{\pythia}{{\tt PYTHIA}}
\newcommand{\mcatnlo}{{\tt MC@NLO}}

\newcommand{\jimmy}{{\tt JIMMY}}

\newcommand{\mct}{\ensuremath{m_\mathrm{CT}}}

\def\lsim{\mathrel{\rlap{\lower4pt\hbox{\hskip1pt$\sim$}}
    \raise1pt\hbox{$<$}}}                
\def\gsim{\mathrel{\rlap{\lower4pt\hbox{\hskip1pt$\sim$}}
    \raise1pt\hbox{$>$}}}                
\makeatletter
\renewcommand{\p@subfigure}{\arabic{figure}}
\renewcommand{\thesubfigure}{\alph{subfigure}}
\renewcommand{\@thesubfigure}{(\thesubfigure)}
\makeatother

\begin{document}

\begin{frontmatter}

\title{\vspace{-0.5cm} \flushright{\normalsize{CERN-PH-EP-2011-165}} \\ \center{Searches for supersymmetry with the ATLAS detector using final states with two leptons and missing transverse momentum in $\sqrt{s}=7$~TeV proton-proton collisions}}
\journal{Physics Letters B}

\author{The ATLAS Collaboration}

\begin{abstract}

Results of three searches are presented for the production of supersymmetric particles decaying into final states with missing transverse momentum and exactly two isolated leptons, $e$ or $\mu$. The analysis uses a data sample collected during the first half of 2011 that corresponds to a total integrated luminosity of 1~fb$^{-1}$ of $\sqrt{s}=7$~TeV proton-proton collisions recorded with the ATLAS detector at the Large Hadron Collider. Opposite-sign and same-sign dilepton events are separately studied, with no deviations from the Standard Model expectation observed. Additionally, in opposite-sign events, a search is made for an excess of same-flavour over different-flavour lepton pairs. Effective production cross sections in excess of 9.9~fb for opposite-sign events containing supersymmetric particles with missing transverse momentum greater than 250~GeV are excluded at 95$\%$ CL. For same-sign events containing supersymmetric particles with missing transverse momentum greater than 100~GeV, effective production cross sections in excess of 14.8~fb are excluded at 95$\%$ CL. The latter limit is interpreted in a simplified electroweak gaugino production model excluding chargino masses up to 200~GeV, under the assumption that slepton decay is dominant.
\end{abstract}
\end{frontmatter}

\section{Introduction}
Many extensions to the Standard Model (SM) predict the existence of new states that decay to invisible particles. New coloured particles, such as the squarks ($\tilde{q})$ and gluinos ($\tilde{g})$ of supersymmetric (SUSY) theories~\cite{susy}, are among those predicted. These new particles could be accessible at the Large Hadron Collider (LHC).  In R-parity conserving~\cite{r-parity} SUSY models, the lightest supersymmetric particle (LSP) is stable and weakly interacting, and SUSY particles are pair-produced. The LSP escapes detection, giving rise to events with significant missing transverse momentum ($E^{\mathrm{miss}}_{\mathrm{T}}$). The dominant SUSY production channels at the LHC are: squark-(anti)squark, squark-gluino and gluino pair production. The squarks and gluinos are expected to decay into quarks and the SUSY partners of the gauge and Higgs bosons, charginos, $\tilde{\chi}^{\pm}$, and neutralinos, $\tilde{\chi}^{0}$. Weak gauginos and sleptons may also be pair-produced, albeit with smaller cross sections, and dilepton searches are potentially very sensitive to direct electroweak gaugino production: $\tilde{\chi}^{\pm}_{1}\tilde{\chi}^{0}_{2}$, $\tilde{\chi}^{\pm}_{2}\tilde{\chi}^{0}_{1}$, $\tilde{\chi}^{\pm}_{1}\tilde{\chi}^{\mp}_{1}$ and $\tilde{\chi}^{0}_{2}\tilde{\chi}^{0}_{2}$.

SUSY events can produce charged leptons with high transverse momentum ($p_\mathrm{T}$) through the decays of neutralinos and charginos. The main processes are: (a)~$\tilde{\chi}^{0}_{i}\rightarrow l^{\pm}\nu\tilde{\chi}^{\mp}_{j}$, (b)~$\tilde{\chi}^{\pm}_{i}\rightarrow l^{\pm}\nu\tilde{\chi}^{0}_{j}$, (c)~$\tilde{\chi}^{0}_{i}\rightarrow l^{\pm} l^{\mp} \tilde{\chi}^{0}_{j}$ and (d)~$\tilde{\chi}^{\pm}_{i}\rightarrow l^{\pm}l^{\mp}\tilde{\chi}^{\pm}_{j}$, where $l$ is an $e$, $\mu$ or $\tau$ lepton (only $e$ and $\mu$ are considered in this paper). These decays can be direct, or proceed via an intermediate slepton.

In each SUSY event there are two independent cascade decays. Two leptons are produced in events in which two gauginos decay via cascade a) or b), or events in which one gaugino
decays via cascade c) or d). In the former case, the events may contain same
sign leptons and the lepton flavour may differ. In the latter case, the
leptons will have the same flavour, and searching for an excess of opposite-sign same-flavour dilepton events over different flavour events offers one of the best
routes to the model-independent measurement of SUSY particle masses
via end-points in the dilepton invariant mass distribution \cite{Hinchliffe:1996iu,Allanach:2000kt,Costanzo:2009mq}.

Previous results of SUSY searches at the LHC for final states with two leptons, electrons or muons, can be found in Refs.~\cite{atlas2lep1,atlas2lep2,cms2lepOS,cms2lepSS}. This letter presents updated results using data recorded during 2011 from each of the three ATLAS searches for SUSY in events with exactly two leptons and significant missing transverse momentum. The two inclusive searches for opposite- and same-sign lepton pairs and the search for an excess of events with same-flavour lepton pairs proceed similarly to those reported in Refs~\cite{atlas2lep1} and~\cite{atlas2lep2}, with minor modifications. The latter is termed a ``flavour subtraction'' analysis, and considers the subtraction of different-flavour dilepton events from those of same-flavour. 
In the 2011 analyses, the rejection criteria for cosmic ray muons are stricter and the method for estimating their contamination to the signal regions is modified. Lepton kinematic selection criteria are also adjusted to match the single lepton triggers used in 2011. The experimental environment differs significantly from that of 2010 due to the higher rate of multiple proton-proton collisions per bunch-crossing (pile-up) produced by the LHC.

In 2010, the dilepton analyses set limits in high-$E^{\mathrm{miss}}_{\mathrm{T}}$ signal regions, $E^{\mathrm{miss}}_{\mathrm{T}}>100 (150)$~GeV for opposite-sign (same-sign) analyses. In this 2011 analysis, a wider variety of signal regions is considered, placing requirements on $E^{\mathrm{miss}}_{\mathrm{T}}$, but also on the number of high-$p_\mathrm{T}$ jets (see Table~\ref{OS_SR}). Additionally, exclusion limits are set in a simplified model of electroweak gaugino production (in these simplified models the LSP is bino-like and the effect of a Higgsino admixture in the chargino and neutralino states not considered). Previous limits on electroweak gaugino production can be found in Refs.~\cite{Abazov:2009zi,Forrest:2009gm,lepsusy,Heister:2001nk,Heister:2003zk,Abdallah:2003xe,Achard:2003ge,Abbiendi:2003ji}. These limits are not directly comparable to those in this paper because of the assumptions made for the simplified models considered.

\section{The ATLAS Detector}

The ATLAS detector~\cite{Aad:2008zzm} is a multi-purpose particle
physics apparatus with a forward-backward symmetric cylindrical
geometry and nearly 4$\pi$ coverage in solid angle.\footnote{
ATLAS uses a right-handed coordinate system with its origin at the nominal
interaction point in the centre of the detector and the $z$-axis along the
beam pipe. Cylindrical coordinates $(r,\phi)$ are used in the transverse
plane, $\phi$ being the azimuthal angle around the beam pipe. The pseudorapidity $\eta$ is
defined in terms of the polar angle $\theta$ by $\eta=-\ln\tan(\theta/2)$.
}  It contains four superconducting magnet systems, which comprise a 
thin solenoid surrounding the inner tracking detector (ID), and barrel and endcap toroids supporting a muon spectrometer. The ID consists of a silicon pixel detector, a silicon microstrip detector (SCT), and a transition
radiation tracker (TRT). The muon spectrometer surrounds the calorimeters and consists of a system of precision tracking chambers ($|\eta|<2.7$), and detectors for triggering ($|\eta|<2.4$).
 In the pseudorapidity
region $\left|\eta\right| < 3.2$, high-granularity liquid-argon (LAr)
electromagnetic (EM) sampling calorimeters are used.  An iron-scintillator tile
calorimeter provides coverage for hadron detection over
$\left|\eta\right| < 1.7$.  The end-cap and forward regions, spanning
$1.5~<~\left|\eta\right|~<~4.9$, are instrumented with LAr calorimetry
for both EM and hadronic measurements.

\section{Trigger and Data Sample}

The data used in this analysis were recorded between March and June 2011, with the LHC operating at a centre-of-mass energy of 7~\TeV.
Application of beam, detector and data-quality requirements gives a total integrated luminosity of 1.04~fb$^{-1}$, with an estimated uncertainty of 3.7$\%$~\cite{lumicorrect}.

Events must pass either a single electron or a single muon trigger. The $p_\mathrm{T}$ thresholds of these triggers are 20~GeV and 18~GeV respectively. These triggers reach full efficiency for electrons with $p_\mathrm{T}>25$~GeV and muons with $p_\mathrm{T}>20$~GeV, with typical efficiencies for leptons selected for offline analysis of 96$\%$ for electrons, and of 75$\%$ and 88$\%$ for muons in the barrel ($|\eta|<1.05$) and end-cap ($1.05~<~|\eta|~<~2.4$) regions, respectively.

\section{Monte Carlo}

\label{sec:MC}

Monte Carlo (MC) simulated event samples are used to develop and validate the analysis procedure and to help evaluate the SM backgrounds in the various signal regions. Production of top quark pairs is simulated with \mcatnlo~\cite{mcatnlolong}, using a top quark mass of $172.5$~GeV and the next-to-leading order (NLO) parton distribution functions (PDF) {\tt CTEQ6.6}~\cite{cteq66}, which are used with all NLO MC codes in this analysis. Samples of $W$ production and Z/$\gamma^*$ production, with accompanying jets, are produced
with \alpgen~\cite{Mangano:2002ea}\footnote{The MC samples for $Z/\gamma^*+$jets are divided into two invariant mass windows. The first cover $10<m_{ll}<40$~GeV and are referred to in this paper as ``Drell-Yan'' events. The second cover the region $m_{ll}>40$~GeV and are referred to as $~Z+$jets}. Diboson ($WW$, $WZ$, $ZZ$) production is simulated
with \herwig~\cite{herwig}, $W^+W^+jj$ production with MadGraph~\cite{madgraph} and single top production 
with \mcatnlo. Fragmentation and hadronisation for the \alpgen~and \mcatnlo~samples are performed with \herwig, using
\jimmy~\cite{Butterworth:1996zw} for the underlying event. \alpgen~and \powheg~\cite{powheg}~samples are used to assess the systematic uncertainties associated with the choice of generator for $t\bar{t}$ production, and {\tt AcerMC}~\cite{Kersevan:2004yg} samples are used to assess the uncertainties associated with initial and final state radiation (ISR/FSR). The simplified electroweak gaugino production models are simulated using {\tt HERWIG++}~\cite{herwigpp}, with cross sections calculated at NLO using {\tt PROSPINO}~\cite{prospino}. Samples of QCD jet events are generated with \pythia~using the {\tt MRST2007LO*} modified leading-order PDF~\cite{Sherstnev:2007nd}, which are used with all leading-order MC codes in this analysis. The QCD jet MC is only used for cross-checks of components of the data-driven background estimation.

The MC samples are produced using the ATLAS MC10b parameter 
tune \cite{mc10a} and a {\tt GEANT4}~\cite{geant4} based detector simulation~\cite{atlassimulation}. 
MC samples are reweighted so that the number of interactions per bunch crossing agrees with that in data.

\section{Object Reconstruction}

\label{sec:objectreconstruction} 

Electrons are reconstructed from clusters in the electromagnetic calorimeter
matched to a track in the ID. Electrons are required to pass the ``medium''~\cite{Aad:2010yt} electron definition (selection criteria based mainly on lateral shower shape requirements in the calorimeter) and have $\pt > 20 \GeV$ and $|\eta|~<~2.47$. Electrons within $0.2 < \Delta R < 0.4$ of any jet are discarded, where
$\Delta R~=~\sqrt{(\Delta \eta)^2 + (\Delta \phi)^2}$. When the jet-electron distance is below 0.2, the jet is removed.
For electrons in the signal region, the quality criterion is raised to ``tight'' by placing additional requirements on the ratio of calorimetric energy to track momentum, and the number of high-threshold hits in the TRT. Furthermore, the electrons are required to be isolated: the $\pt$ sum of
tracks above 1~GeV within a cone of size $\Delta R < 0.2$ around each electron candidate
(excluding the electron candidates themselves) is required to be less than
10\% of the electron's $\pt$. If the electron is the highest $p_\mathrm{T}$ lepton in the pair, the $\pt$ requirement is raised to
$25 \GeV$.

Muons are reconstructed using either a full muon spectrometer track matched to an ID track, or a muon spectrometer track segment matched to an extrapolated ID track.
Muons are required to have $\pt > 10 \GeV$, $|\eta| < 2.4$,
and to be well reconstructed, with sufficient hits in the pixel, SCT, and TRT detectors.
Muon tracks reconstructed independently in both the ID and muon spectrometer are required to have a good match and a compatible momentum measurement in both detectors.
Muons within $\Delta R < 0.4$ of any jet are discarded. In order to reject muons resulting from cosmic rays, tight
cuts are applied to the origin of the muon relative to the
primary vertex (PV): muon tracks are required to have a longitudinal impact parameter $|z_0| < 1$~mm and a transverse impact parameter $|d_0| < 0.2$ mm. 
Muons in the signal region must be isolated: the $\pt$ sum
of tracks within a cone of size $\Delta R < 0.2$ around the muon candidate
(excluding the muon candidate itself) is required to be less than $1.8 \GeV$.
If a muon in a signal region is the highest $p_\mathrm{T}$ lepton in the pair, the $\pt$ requirement is raised to
$20 \GeV$.

Jets are reconstructed using the anti-$k_t$ jet clustering 
algorithm~\cite{antikt} with a distance parameter of 0.4. 
The inputs to the jet algorithm are clusters formed
from energy deposits in the calorimeter.
Jets are required to have $\pt > 20 \GeV$
and $|\eta| < 2.8$. Events with any jet that fails quality criteria designed to remove noise and non-collision backgrounds~\cite{atlas-jet-clean} are rejected.

The missing transverse momentum ($\met$) in this analysis is the magnitude of the vector sum of the $\pt$ of reconstructed objects in the event. The objects considered are jets
with $\pt > 20 \GeV$, signal leptons, any additional non-isolated
muons (for example from semi-leptonic decays of hadrons in jets) and calorimeter clusters with $|\eta|<4.5$ which are not associated to any of the aforementioned objects.

\section{Event Selection}\label{sec:eventselection}

The primary vertex (the vertex with the highest summed track $p_\mathrm{T}^2$) in each event is required to have at least five associated tracks. Due to 
readout problems in the LAr calorimeter for a subset of the data, events in data and MC containing a jet with $p_\mathrm{T}>20$~GeV or an identified electron with $-0.1<\eta<1.5$ and $-0.9<\phi<-0.5$ are rejected (resulting in a loss of less than 2\% of the data).
Each selected event must contain exactly two reconstructed leptons, $e$ or $\mu$, satisfying the conditions described in Section~\ref{sec:objectreconstruction}. Events containing exactly two electrons (muons) must satisfy the electron (muon) trigger. For events containing exactly one electron and one muon: those with an electron with $p_\mathrm{T}>25$~GeV must satisfy the electron trigger, while events with no such electron must have a muon with $p_\mathrm{T}>20$~GeV and satisfy the muon trigger. Events containing an electron with $p_\mathrm{T}>25$~GeV which do not satisfy the electron trigger are recovered using the muon trigger provided the $p_\mathrm{T}$ of the muon is greater than 20~GeV.

Additionally, both leptons in each pair must satisfy the signal region requirements. To remove low-mass dilepton resonances, the invariant mass ($m_{ll}$) of the lepton-pair must be greater than 12~GeV. The selected events are then classified as opposite-sign or same-sign, depending on the respective charges of each lepton in the pair.

The various signal regions defined for the opposite-sign (OS-x), same-sign (SS-x) and flavour-subtraction (FS-x) analyses are given in Table~\ref{OS_SR}. The opposite-sign and same-sign signal regions are designed to provide sensitivity to R-parity conserving SUSY models with high-$E^{\mathrm{miss}}_{\mathrm{T}}$ (OS-inc and SS-inc) and electroweak gaugino production (SS-inc). Signal regions that introduce requirements on the multiplicity and $p_\mathrm{T}$ of jets in the events (OS-3j, OS-4j and SS-2j) exploit the expected presence of jets in 
cascade decays from coloured SUSY particle production. The three latter regions 
are optimised by considering their potential reach in the parameter space of mSUGRA/CMSSM~\cite{susy} models\footnote{These models have varying universal scalar and gaugino mass parameters $m_0$ and $m_{1/2}$, but fixed values of the universal trilinear coupling parameter $A_0=0$~GeV, ratio of the vacuum expectation values of the two Higgs doublets $\tan\beta=10$, and Higgs mixing parameter, $\mu>0$.}. For the flavour-subtraction analysis, the signal regions aim to fully exploit the natural cancellation of $t\bar{t}$ and other flavour-symmetric background events and to have a minimum contamination from $Z/\gamma^*+$jets and diboson events. The contamination from flavour-asymmetric background is reduced with either a veto on events with $m_{ll}$ near the mass of the $Z$ boson (FS-no Z), requirements on jet multiplicity and $p_\mathrm{T}$ (FS-2j) or very high-$E^{\mathrm{miss}}_{\mathrm{T}}$ (FS-inc).
\begin{table*}[!htbp]
\begin{center}
\begin{tabular}{l|r|r|r|r|r|r|r|r}
\hline
Signal Region & OS-inc & OS-3j & OS-4j & SS-inc & SS-2j & FS-no Z & FS-2j & FS-inc\\
\hline
\hline
$E^{\mathrm{miss}}_{\mathrm{T}}$ [GeV] & 250 & 220 & 100 & 100 & 80 & 80 & 80 & 250\\
\hline
Leading jet $p_\mathrm{T}$ [GeV] & - & 80 & 100 & - & 50 & - & - & -\\
Second jet $p_\mathrm{T}$ [GeV] & - & 40 & 70 & - & 50 & - & - & -\\
Third jet $p_\mathrm{T}$ [GeV] & - & 40 & 70 & - & - & - & - & -\\
Fourth jet $p_\mathrm{T}$ [GeV] & - & - & 70 & - & -& - & - & -\\
\hline
Number of jets & - & $\geq3$ & $\geq4$ & - & $\geq2$ & - & $\geq2$ & - \\
$m_{ll}$ veto [GeV] & - & - & - &  - & - & 80-100 & -& -\\
\hline
\end{tabular}
\end{center}
\caption{
\label{OS_SR} Criteria defining each of the three signal regions for the opposite-sign (OS-x) analysis, each of the two signal regions for the same-sign analysis (SS-x) and each of the three regions for the flavour-subtraction (FS-x) analysis. Regions OS-inc and FS-inc are identical.}
\end{table*}

\section{Background Evaluation}\label{sec:backgrounds}

The background from cosmic rays must be evaluated in all signal regions.
Muons from hard scattering processes typically have very low values of $|z_0|$ and $|d_0|$ since they originate from the PV of the event. The distributions of both $|z_0|$ and $|d_0|$ for cosmic rays are broad. In the $\mu\mu$ channels the expected numbers of cosmic ray events in each signal region are evaluated using the $|z_0|$ distribution of muons in dimuon events for which the $|z_0|$ and $|d_0|$ requirements have been relaxed. The region $1<|z_0|<100$~mm is populated with cosmic rays. Due to the fall off of the tracking efficiency at large $z_0$, this region can be well described by a Gaussian fit. This fit can be used to evaluate the number of cosmic rays in the region $|z_0|<1$~mm, given the estimated number in the region $1<|z_0|<100$~mm after the application of the signal region selection cuts. 
This procedure yields contributions from cosmic rays of $< 10^{-3}$ events in each signal region.
The coincidence of a single reconstructed collision electron and a single reconstructed cosmic ray muon is much less likely than the probability of reconstructing a cosmic ray event as two reconstructed muons in coincidence with a collision event. This sets a conservative estimate of the contribution in the $e\mu$ channels of $< 10^{-3}$ events. 

The SM backgrounds to each search are evaluated using a combination of MC simulation and data-driven techniques. Contributions from single top and diboson events are evaluated using the MC samples described in Section~\ref{sec:MC}, scaled to the luminosity of the data sample. The former must be evaluated only in OS-x and FS-x signal regions, while the latter must be evaluated in all signal regions.
Contributions from $Z/\gamma^*+$jets and $t\bar{t}$ events (which must be estimated in OS-x and FS-x signal regions, but not SS-x regions) are evaluated using MC samples normalised to data in appropriate control regions (CR). SM processes generating events containing at least one fake or non-isolated lepton are collectively referred to as ``fake lepton'' background, generally consisting of semi-leptonic $t\bar{t}$, single top, $W+$jets and QCD light and heavy-flavour jet production. The fake lepton background is obtained using a purely data-driven technique for all signal regions. The background from charge misidentification (from electrons in events which have undergone hard bremsstrahlung with subsequent photon conversion) is important in the same-sign signal region and is estimated using a partially data-driven technique.

The following paragraphs first describe the evaluation of the backgrounds which contribute only to the opposite-sign (and flavour-subtraction) signal regions. The fake lepton background for all signal regions is then described. Lastly, details are given of how the background from charge misidentification is estimated for each same-sign signal region.

The fully leptonic $t\bar{t}$ background in the signal regions is obtained by extrapolating from the number of $t\bar{t}$ events in a suitable control region, after correcting for contamination from non-$t\bar{t}$ events, into the signal regions using the ratio of the number of MC $t\bar{t}$ events in the signal region to those in the control region.
The numbers of $t\bar{t}$ events in a given control region 
are determined using a ``top-tagging'' algorithm.
The top-tagging requirement is imposed through the use of the variable $\mct$ \cite{Tovey:2008}. This observable can be calculated from the four-vectors of the selected jets and leptons:
\begin{align}
  \mct^{2}(v_{1},v_{2}) = \left[ E_{\rm T}(v_{1}) + E_{\rm T}(v_{2}) \right]^2 \notag\\
  -  \left[  {\bf p_{\rm T}}(v_{1}) - {\bf p_{\rm T}}(v_{2}) \right]^2 ,
  \label{eq:mct_endpoints}
\end{align}
where $v_i$ can be a lepton ($l$), a jet ($j$), or a lepton-jet combination ($jl$), transverse momentum vectors are defined by ${\bf p_{\rm T}}$ and transverse energies $E_\mathrm{T}$ are defined as $E_\mathrm{T} = \sqrt{p_{\rm T}^2 +m^2}$. The quantities $m_{\textup{CT}}(j,j)$, $m_{\textup{CT}}(l,l)$ and $m_{\textup{CT}}(jl,jl)$ are bounded from above by analytical functions of the top quark and $W$ boson masses. A top-tagged event must have at least two jets with $p_\mathrm{T}>20$~GeV, and the scalar sum of the $p_\mathrm{T}$ of at least one combination of two jets and the two leptons in the event must exceed 100~GeV. Furthermore, top-tagged events are required to possess $\mct$ values calculated from combinations of jets and leptons consistent with the expected bounds from $t\bar{t}$ events as described in Ref.~\cite{Polesello:2009rn} ($m_\textup{CT}(jj)$ in the allowed area of the $m_\textup{CT}(jj)$-$p_T(jj)$ plane, $m_\textup{CT}(l_1,l_2)$ in the allowed area of the $m_\textup{CT}(l_1,l_2)$-$p_T(ll)$ plane and $m_\textup{CT}(jl,jl)$ compatible with $\tbar{t}$) as well as lepton-jet invariant mass values consistent with top quark decays ($m(j_1l_1)<155$~GeV and $m(j_1l_2)<155$~GeV). 
The contributions in each opposite-sign signal region are obtained using three separate control regions (one for each signal region).
All three control regions (for OS/FS-inc, OS-3j and OS-4j) require, in addition to the top-tagged lepton pairs, $60~<~E^{\mathrm{miss}}_{\mathrm{T}}~<~100$~GeV, except in the $e^\pm e^\mp$ and $\mu^\pm\mu^\mp$ channels of OS-inc, where $80<E^{\mathrm{miss}}_{\mathrm{T}}<100$~GeV is required. In the first (a control region for OS/FS-inc), no requirement is placed on the jets, while in the second (for OS-3j) and third (for OS-4j), three jets and four jets with $p_\mathrm{T}>40$~GeV are required respectively. In these control regions the numbers of observed events (1010, 238 and 52 in control regions one through to three, respectively) are in good agreement (better than 1 $\sigma$) given statistical and systematic uncertainties with the expected rates from $t\bar{t}$ and non-$t\bar{t}$ SM processes, resulting in ratios of data to MC in the control regions compatible with one. The contamination from non-$t\bar{t}$ events lies between 15 and 20$\%$. In the first two signal regions for the flavour subtraction analysis (FS-no Z and FS-2j), the contribution from fully-leptonic $t\bar{t}$ is taken from MC.

Similarly, the contribution from $Z/\gamma^*+$jets events in the signal regions is estimated by extrapolating the number of $Z/\gamma^*+$jets events observed in a control region into the signal region using ratios derived from MC.
All $Z/\gamma^*$ control regions contain lepton-pair events satisfying the same selection criteria as the signal region but with $\MET<20$~GeV and an additional $81~<~m_{\ell\ell}~<~101$~GeV requirement. Three distinct control regions are necessary for the three different opposite-sign signal regions: the first (a control region for OS/FS-inc) places no requirements on the number of jets in the event, while the second and third (for OS-3j and OS-4j respectively) require jets with $p_\mathrm{T}$ as described in Table~\ref{OS_SR}. Similarly, in the control regions for the flavour-subtraction signal regions (FS-no Z and FS-2j), the corresponding jet requirements in Table~\ref{OS_SR} are used. In these control regions the numbers of observed events are in good agreement with the expectation from MC, given the systematic and statistical uncertainties on the MC expectation. The predicted numbers of $Z/\gamma^*+$jets events in each signal region are compatible with the MC expectation (within 1 $\sigma$).

The probabilities of fake leptons being reconstructed as prompt, isolated leptons are evaluated from suitable control regions. 
Putative fake leptons are identified as those satisfying a \emph{loose} set of identification requirements, and the fraction of these that pass the \emph{tight} identification requirements used for signal leptons is measured.
For muons, the looser identification requirements are identical to those of the signal muons, except the isolation requirement is dropped. Looser electrons must be both ``medium''  as defined in Ref.~\cite{Aad:2010yt} and not isolated, but are otherwise identical to the signal electrons. The probability of identifying a heavy-flavour decay, light-jet or photon conversion as a prompt electron is evaluated from events with a single electron satisfying the relaxed identification requirements, $E^{\mathrm{miss}}_{\mathrm{T}}<30$~GeV, at least one jet and $\Delta\phi$ between the lepton and $E^{\mathrm{miss}}_{\mathrm{T}}$ directions less than 0.5 (reducing $W$ backgrounds). The corresponding control region for estimating the prompt muon misidentification probability also requires $E^{\mathrm{miss}}_{\mathrm{T}}<30$~GeV and selects events with two same-sign muons satisfying the relaxed identification requirements. The contamination from processes producing prompt, isolated leptons has been studied in MC simulations and is small. With this ``lepton'' definition, both control regions and signal regions have a similar composition and are dominated by heavy-flavour decays, light-jets, and photon conversions. In each signal region (OS-x, SS-x and FS-x) the observed numbers of events in data with two loose leptons, two tight leptons, or one of each are counted. Systems of linear equations are then constructed for each signal region relating the observed numbers of events with two fake leptons, two real leptons, or one of each to the observed event counts, using the measured probabilities of misidentification for fakes and efficiencies for identification of real leptons. The latter are obtained for electrons and muons separately from events with a single same-flavour opposite-sign lepton pair with $m_{ll}$ within 5~GeV of the $Z$ mass. Simultaneous solution of these equations in each signal region yields the expected number of events in each which contain fake leptons. This method is the ``matrix-method'' described in Ref.~\cite{atlas-topxsec}.

The contribution from charge misidentification in each SS-x region is studied using $Z\rightarrow e^+e^-$ MC events. The probability of charge misidentification is ascertained by comparing the charges of generator level electrons to those of reconstructed electron candidates following the application of the same-sign signal region cuts. The misidentification probability is calculated as a function of electron rapidity and transverse momentum and applied to $t\bar{t}\rightarrow e^\pm l^\mp$ ($l=e,\mu$) MC events to evaluate, in each signal region, the number of same-sign events from incorrect charge assignment. The charge misidentification probabilities in the $Z\rightarrow e^+e^-$ and $t\bar{t}$ MC samples are consistent. A single scaling factor is used to correct for discrepancies between the charge misidentification rates in data and simulation. The $p_\mathrm{T}$ distributions in data and MC are in good agreement. The probability of misidentifying the charge of a muon and the contributions from charge misidentification of $Z/\gamma^*+$jets and other SM backgrounds are negligible.

\begin{figure*}[htp]
\begin{center}
\subfigure[]{\includegraphics[scale=0.45]{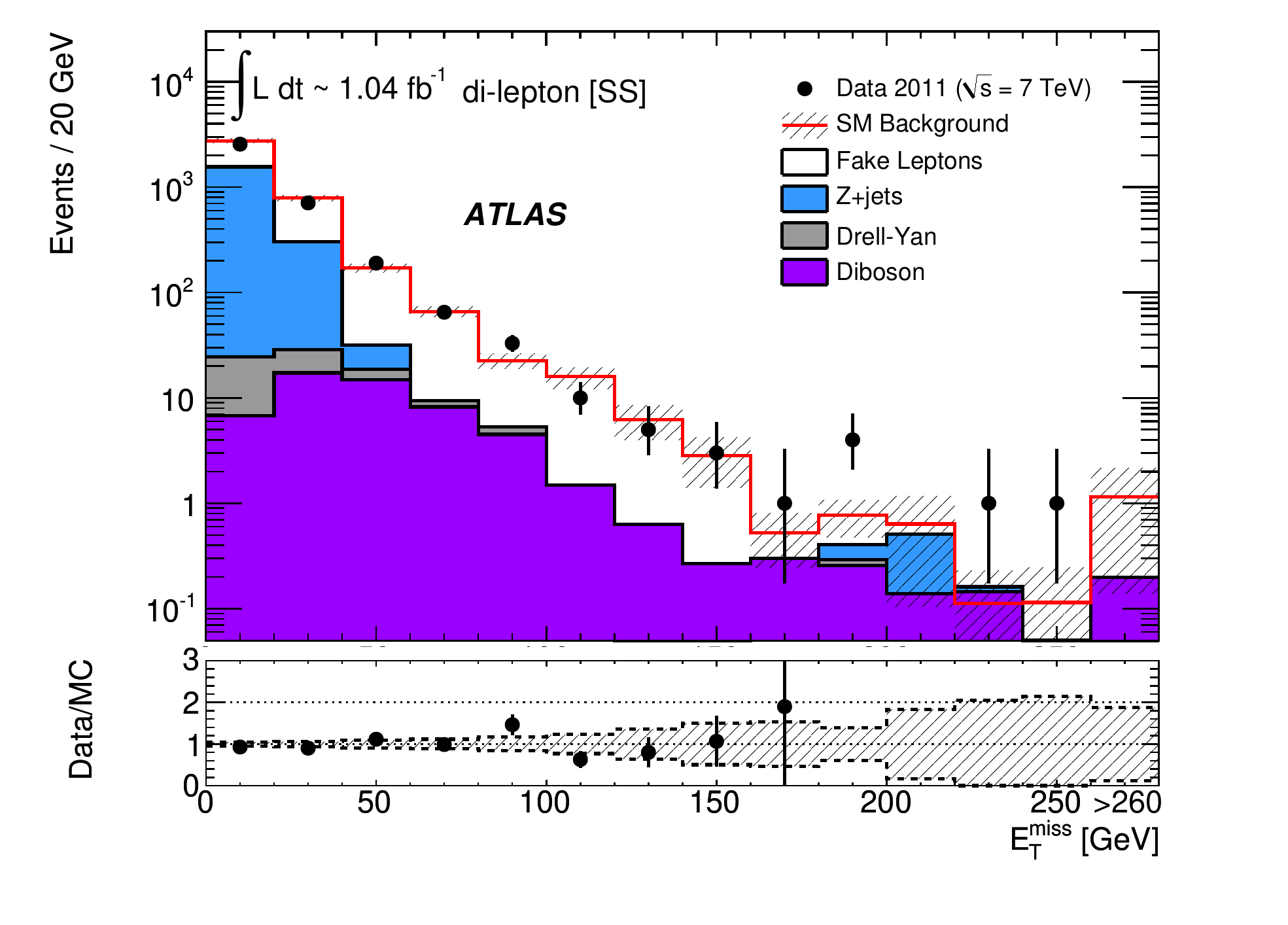}}
\subfigure[]{\includegraphics[scale=0.45]{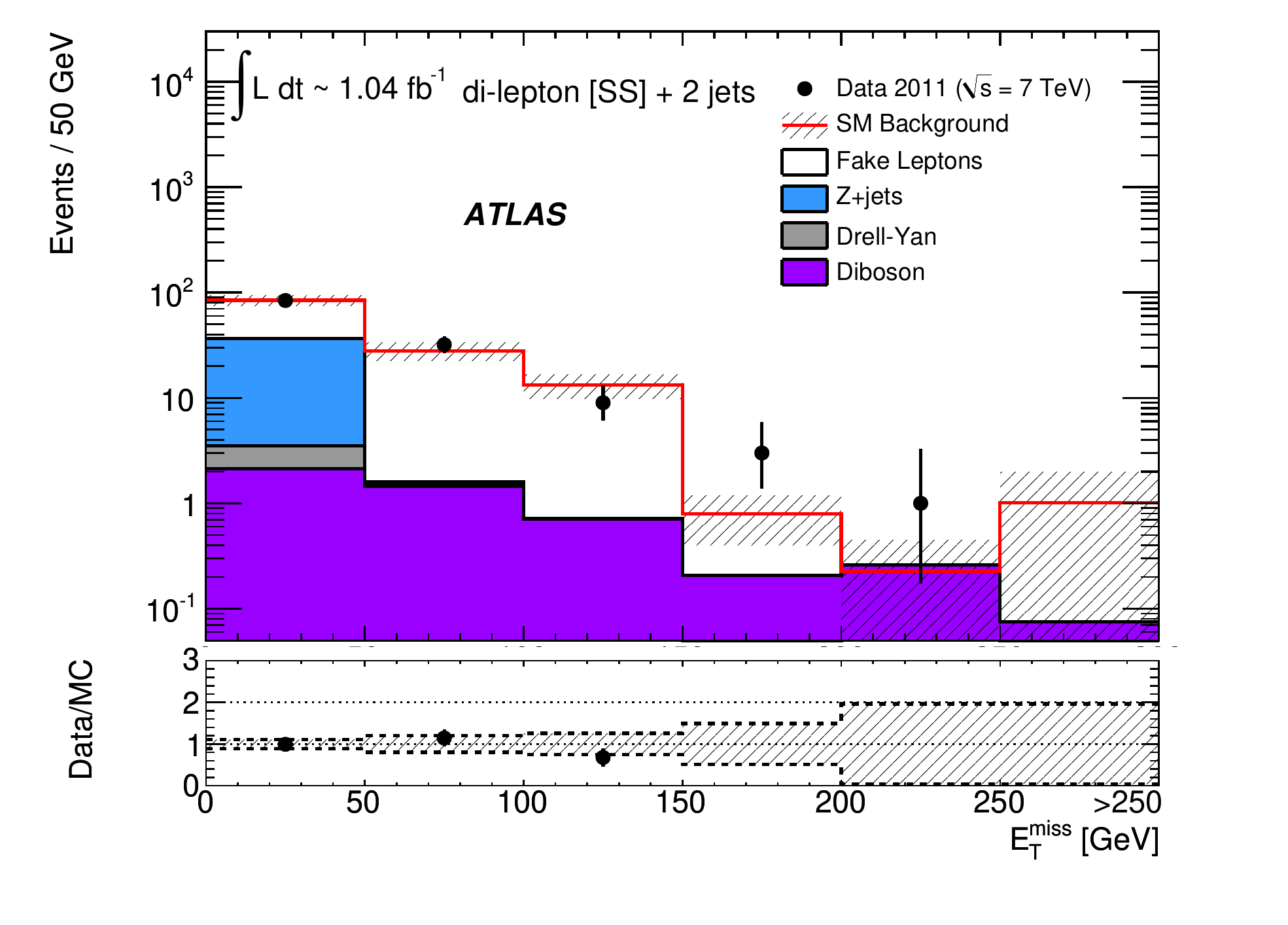}}\newline
\subfigure[]{\includegraphics[scale=0.45]{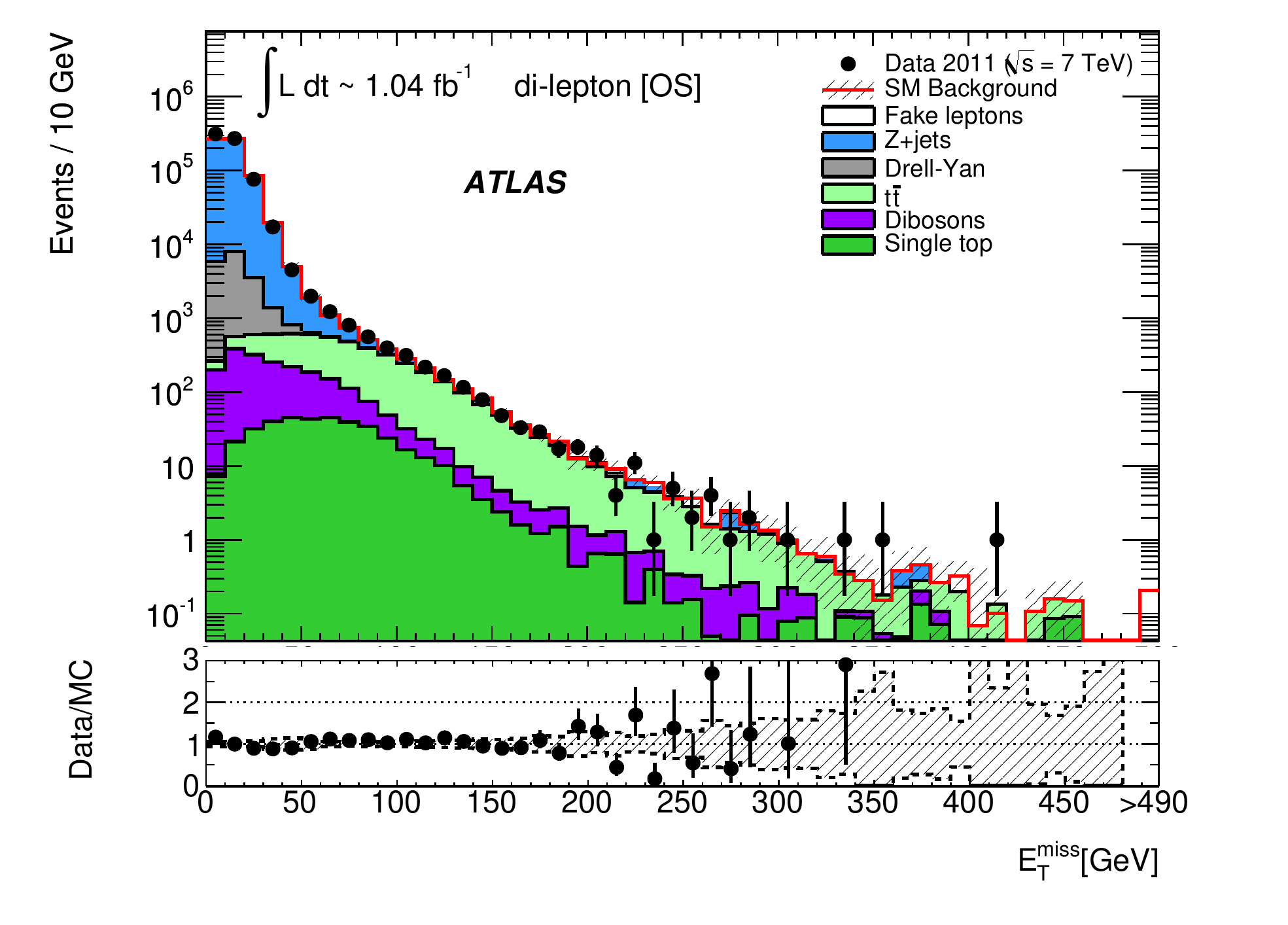}}
\subfigure[]{\includegraphics[scale=0.45]{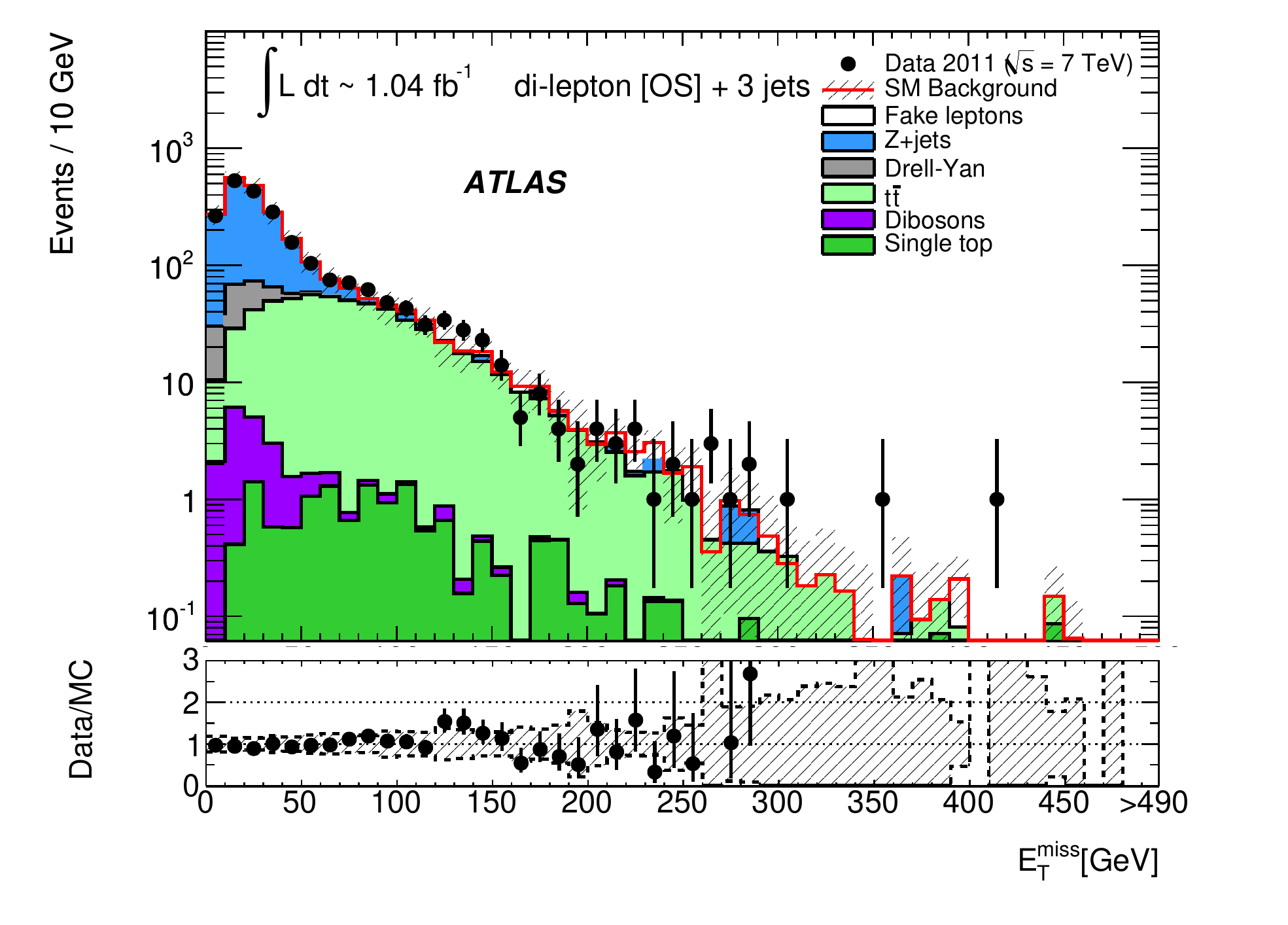}}\newline
\subfigure[]{\includegraphics[scale=0.45]{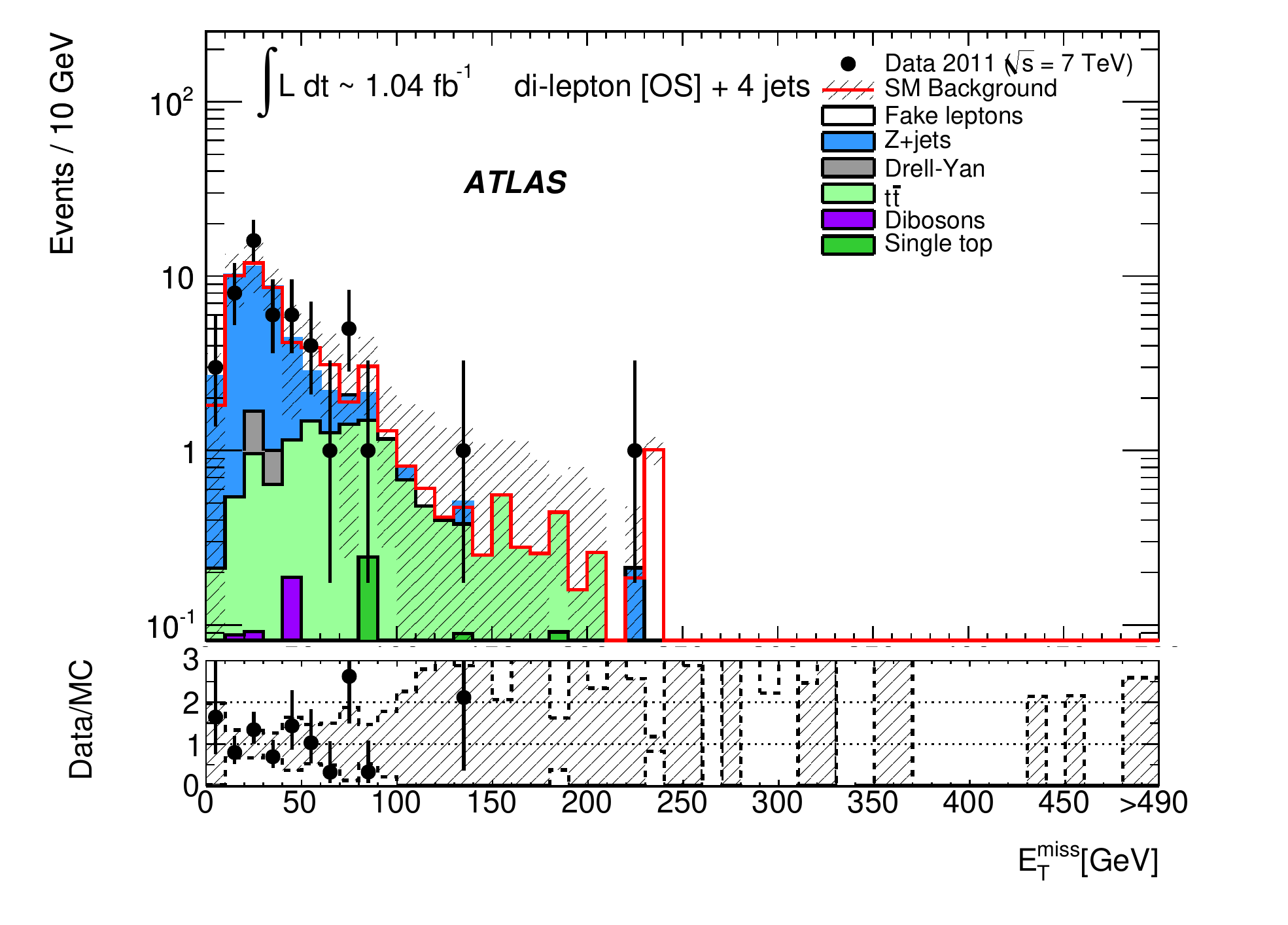}}
\caption{The $E^{\mathrm{miss}}_{\mathrm{T}}$ distributions of same-sign dilepton events before any jet requirement~(a), and after requiring two high-$p_\mathrm{T}$ jets (b) and the $E^{\mathrm{miss}}_{\mathrm{T}}$ distributions of all opposite-sign dilepton events before any jet requirement~(c), after requiring 3 high-$p_\mathrm{T}$ jets~(d) and after the 4 jet requirement~(e). Errors on data points are statistical, while the error band on the SM background represents the total uncertainty. The lower inserts show the ratio between the data and the SM expectation. The component labelled ``Fake leptons'' is evaluated using data as described in the text. The remaining background contributions are from MC, normalised to their respective cross sections and the luminosity of the data sample.  \label{ETmiss_distributions}
}
\end{center}
\end{figure*}

Fully-leptonic $t\bar{t}$ production is the dominant SM background in the search for opposite-sign dileptons, making up at least 50$\%$ of the total SM event yield. Smaller contributions arise from $Z/\gamma^*+$jets, diboson and single-top production, and events containing at least one fake or non-prompt lepton. In all but the highest jet multiplicity opposite-sign signal regions, $Z/\gamma^*+$jets events are the next most significant contribution. After flavour subtraction, flavour-symmetric backgrounds like $t\bar{t}$ naturally cancel.
Events with a fake lepton dominate the same-sign signal samples. Other significant backgrounds come from diboson production and charge mismeasurements. The estimate of the diboson background includes the process $W^+W^+jj$, but neglects $ttW$ which has been found to be insignificant. The relative size of each SM background component in each signal region is illustrated in Fig.~\ref{ETmiss_distributions}. 
\section{Systematic Uncertainties}
\label{sys}

The primary sources of systematic uncertainty on the background event estimations are: the jet energy scale (JES), the jet energy resolution (JER) and theory and MC modelling.  Uncertainties in lepton reconstruction and identification (momentum and energy scales, resolutions and efficiencies) give smaller contributions. The JES and JER uncertainties are jet $p_\mathrm{T}$ and $\eta$ dependent. They are measured using the complete 2010 dataset using the techniques described in Ref.~\cite{jes}, with an additional contribution (7$\%$) added to the JES uncertainty to account for the effect of higher pile-up in the 2011. Theoretical and MC modelling uncertainties are determined by using different generators and varying the amount of ISR/FSR (for $t\bar{t}$), as described in Section~\ref{sec:MC}. Additional uncertainties arise from limited MC statistics. An uncertainty on the luminosity of 3.7$\%$ is included~\cite{lumicorrect}. 

The main systematic uncertainties on the $t\bar{t}$ background in each OS-x region are summarised in Table~\ref{tab:miTopSyst}. The largest uncertainties (generator and ISR/FSR) affect only the scale factor relating the number of MC $t\bar{t}$ events in the control region to the signal region. Since $t\bar{t}$ dominates the event yields in these regions, these uncertainties make up most of the total systematic uncertainty on the estimated opposite-sign background. For the evaluation of the (smaller) contributions from $Z/\gamma^*+$jets events, a large statistical uncertainty on the MC predictions in the control regions dominates the error. The uncertainties on the single top (in OS-x and FS-x) and diboson (in OS-x, SS-x and FS-x) backgrounds are dominated by the JES and JER contributions. The uncertainties on the yields in all signal regions from events containing fake leptons are dominated by the knowledge of the mis-identification probabilities. This uncertainty makes up most of the total uncertainty on the background yields in SS-x. 

\begin{table}[htb!]
\begin{center}
\begin{tabular}{l | c | c | c }  
\hline
Signal Region          & OS-inc & OS-3j & OS-4j \\
\hline
MC \& CR statistics      &  7\% &  10\% & 21\% \\
JES           & 11\% &  6\% &  6\% \\ 
JER               &  1\% & 11\% & 15\% \\ 
Generator                        & 16\% & 13\% & 58\% \\
ISR/FSR                          & 20\% & 16\% & 26\% \\ 
\hline
Total                 & 27\% & 25\% & 68\% \\
\hline
\end{tabular} 

\caption{A summary of the dominant systematic uncertainties on the estimates of the fully-leptonic $t \bar t$ event yields in each opposite-sign signal region. The uncertainties are different in each signal region, because each has a different control region.}
\label{tab:miTopSyst}

\end{center}
\end{table}

Systematic uncertainties on the signal expectations are evaluated through variations of the factorisation and renormalisation scales in {\tt PROSPINO} between half and twice their default values, and by including the uncertainty on $\alpha_s$ and on the PDF provided by {\tt CTEQ6}. Uncertainties are calculated for individual SUSY processes. In the relevant regions of the illustrated mass plane the resulting uncertainties on the signal cross sections are typically 4-8$\%$. Further uncertainties on the numbers of predicted signal events arise from the JES uncertainty (1-18$\%$), luminosity (3.7$\%$) and finite statistics of the signal Monte Carlo samples.

\section{Results and Interpretation}

\subsection{Opposite and Same-sign Inclusive}
The expected and observed numbers of opposite-sign and same-sign lepton-pair
events in each signal region are compared in Table~\ref{Tab:Interpretation} to the background expectation.Ê
Good agreement is observed.
These results are used to set limits on the effective production cross section, the product of the cross section for new phenomena, the kinematic and geometrical acceptance and reconstruction and event selection efficiencies.
Limits are set using the CL$_{s}$ prescription, as described in Ref.~\cite{Read:2002hq}, and setting the upper limit on the effective production cross section as the limit on the number of observed signal events divided by the integrated luminosity. The results are given in Table~\ref{Tab:Interpretation} in each signal region.

\begin{table}[!htbp]
\begin{center}
\begin{tabular}{c|c|c|c}
\hline
& Background & Obs. & 95$\%$ CL\\ \hline
\hline
OS-inc & 15.5 $\pm$ 4.0 & 13 & 9.9~fb\\
OS-3j & 13.0 $\pm$ 4.0 & 17 & 14.4~fb\\
OS-4j & 5.7 $\pm$ 3.6 & 2 & 6.4~fb\\
\hline
SS-inc & 32.6 $\pm$ 7.9 & 25 & 14.8~fb\\
SS-2j & 24.9 $\pm$ 5.9 & 28 & 17.7~fb\\
\hline
\end{tabular}
\end{center}
\caption{
\label{Tab:Interpretation}Predicted number of background events, observed number of events and the corresponding 95$\%$ CL upper limit on A$\times\epsilon\times\sigma$, calculated using the CL$_{s}$ technique, for each opposite-sign and same-sign signal region.}
\end{table}

\begin{figure}[tbp]
\begin{center}
\includegraphics[scale=0.45]{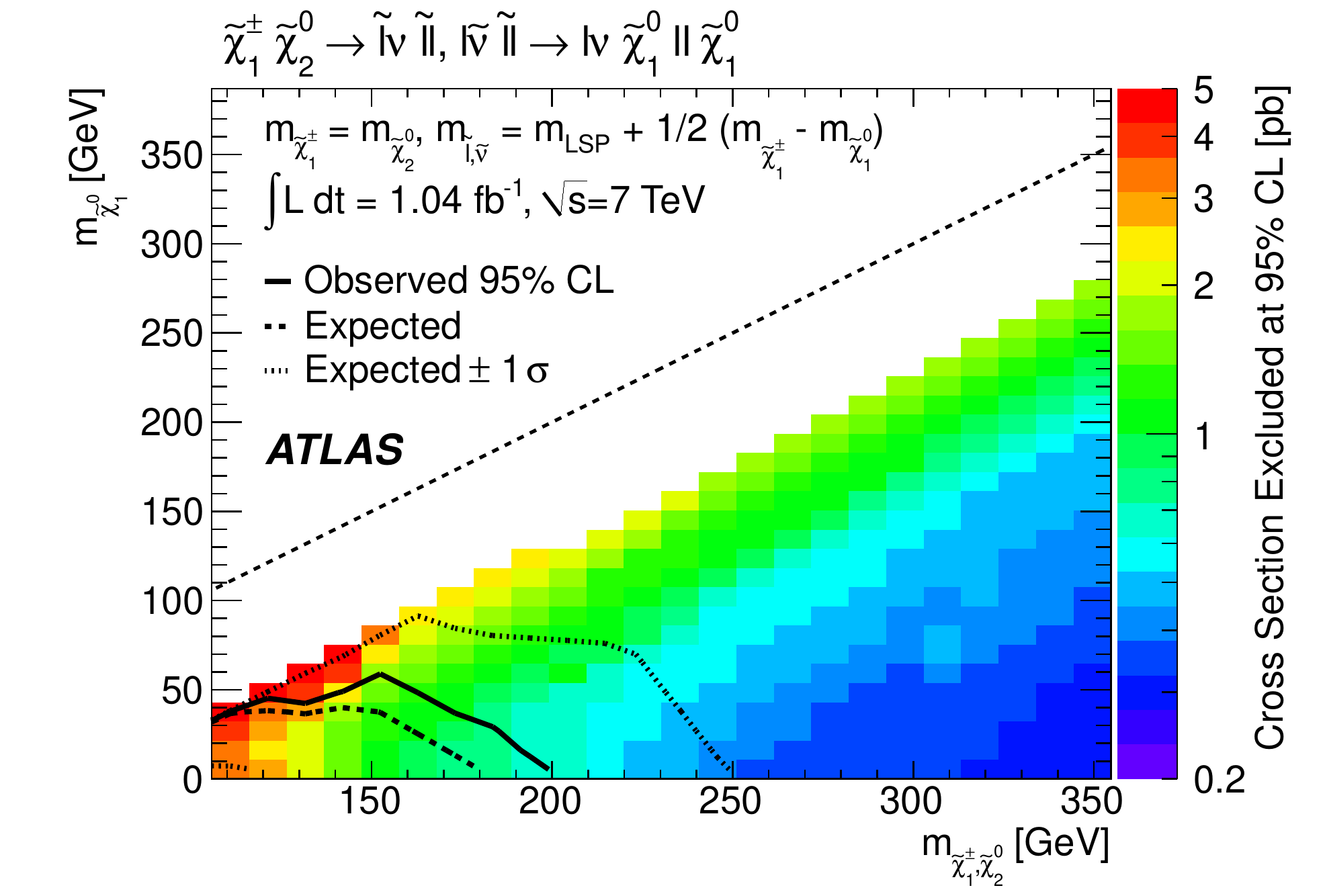}

\caption{95$\%$ CL cross section upper limits (CL$_{s}$) in pb and observed and expected limit contours for $\tilde{\chi}^{\pm}_{1}\tilde{\chi}^{0}_{2}$ production in direct gaugino simplified models.\label{chargino}}
\end{center}
\end{figure}

The signal region SS-inc is particularly sensitive to low mass electroweak gaugino production and the cascade decays into leptons, so only this region is used to set upper limits on the cross section for $\tilde{\chi}^{\pm}_{1}\tilde{\chi}^{0}_{2}$ pair production. The cross section upper limits on $\tilde{\chi}^{\pm}_{1}\tilde{\chi}^{0}_{2}$ pair production, in the simplified direct electroweak gaugino production models detailed in Ref.~\cite{simplified} (Section V, I), are illustrated in Fig.~\ref{chargino} as a function of the $\tilde{\chi}^{\pm}_{1}$ and LSP ($\tilde{\chi}^{0}_{1}$) masses. In this figure, the limits on the effective cross section (taking into account the uncertainties on the signal described in Section~\ref{sys}) are divided by the product of the acceptance and efficiency for each point individually to obtain a grid of limits on the cross section (multiplied by branching ratio). Also shown are the observed and expected limit contours.
The results in Fig.~\ref{chargino} are for slepton masses between the LSP and second lightest neutralino masses and the hierarchy $m_{\tilde{l}}=m_{\tilde{\chi}^{0}_{1}}+\frac{1}{2}(m_{\tilde{\chi}^{\pm}_{1}}-m_{\tilde{\chi}^{0}_{1}})$ with $m(\tilde{\chi}^{\pm}_{1})=m(\tilde{\chi}^{0}_{2})$.

In these simplified models, the squarks are very heavy (permitting only direct $\tilde{\chi}^\pm_1\tilde{\chi}^0_2$ production), the masses of slepton of different flavours are assumed to be degenerate and the branching ratios for both $\tilde{\chi}^{\pm}_{1}\rightarrow\tilde{l}^{\pm}\nu,\tilde{\nu}l^\pm$ and $\tilde{\chi}^{0}_{2}\rightarrow\tilde{l}^{\pm}l^{\mp}$ decays are set to one (with branching ratios for $(\tilde\chi^\pm_1\rightarrow\tilde{l}\nu)$ and $(\tilde{\chi}^\pm_1\rightarrow\tilde{\nu}l)$ equal to 50\%). Furthermore, the sleptons have equal contributions of $\tilde{l}_{L}$ and $\tilde{l}_{R}$, including all slepton and sneutrino flavours. The branching ratio for $(\tilde{l}\rightarrow l\tilde{\chi}^0_1)$ is 100\% and the branching ratio for $(\tilde{\nu}\rightarrow\nu\tilde{\chi}^\pm_1)$ 100\%. 
In this channel, leptons are produced in the cascades: $\tilde{\chi}^{\pm}_{1}\tilde{\chi}^{0}_{2}\rightarrow(\nu \tilde{l}^\pm)(l^\pm\tilde{l}^{\mp})\rightarrow(\nu l^\pm\tilde{\chi}^{0}_{1})(l^\pm l^\mp\tilde{\chi}^{0}_{1})$ and $\tilde{\chi}^{\pm}_{1}\tilde{\chi}^{0}_{2}\rightarrow(l^\pm\tilde{\nu})(l^\pm\tilde{l}^{\mp})\rightarrow(l^\pm\nu\tilde{\chi}^{0}_{1})(l^\pm l^\mp\tilde{\chi}^{0}_{1})$ (with equal branching ratios). The cross section for the line with $m(\tilde{\chi}^{\pm}_{1})=m(\tilde{\chi}^{0}_{2})=200$~GeV is 0.51~pb. Models in the low-mass region have acceptances of $\sim$5-15$\%$ for $\tilde{\chi}^0_1$-$\tilde{\chi}^\pm_{1}$ mass differences from 50 to 200~GeV, and efficiencies of $\sim$20$\%$. If decays to sleptons are dominant, charginos with masses up to 200~GeV are excluded, under the assumptions of these simplified models.

\subsection{Flavour-subtraction Analysis}

In the flavour-subtraction analysis, limits are set on the excess in the number of opposite-sign same-flavour events (multiplied by detector acceptances and efficiencies) in the appropriate signal regions. This is done using pseudo-experiments. The opposite-sign same-flavour excess is quantified using the quantity $\mathcal{S}$, defined as
\small
\begin{align}
\mathcal{S} = \frac{N(e^\pm e^\mp)}{\beta(1-(1-\tau_e)^2)}+\frac{\beta N(\mu^\pm \mu^\mp)}{(1-(1-\tau_\mu)^2)} \notag\\
- \frac{N(e^\pm \mu^\mp)}{1-(1-\tau_e)(1-\tau_\mu)} ,
\label{eq:FlavSubtraction}
\end{align}
\normalsize
which measures the excess of opposite-sign same-flavour events (first two terms) over different-flavour events (third term), taking into account the ratio of electron to muon efficiency times acceptance ($\beta$), and the electron and muon trigger efficiencies ($\tau_e$ and $\tau_{\mu}$), under the assumption that the trigger selection adopted for $e^{\pm}\mu^{\mp}$ events is equivalent to a logical OR of the electron and muon triggers.
This quantity, $\mathcal{S}$ is effectively the excess in the number of same-flavour events multiplied by detector acceptances and efficiencies. The ratio of acceptances and efficiencies, $\beta$, is determined from data to be 0.75$\pm$0.05, with the quoted error including both systematic and statistical uncertainties. The muon trigger efficiency, $\tau_\mu$, averaged over the barrel and end-cap is taken to be (81.6$\pm$ 0.3)$\%$.

\begin{figure*}[ht]
\begin{center}
\subfigure{\includegraphics[scale=0.42]{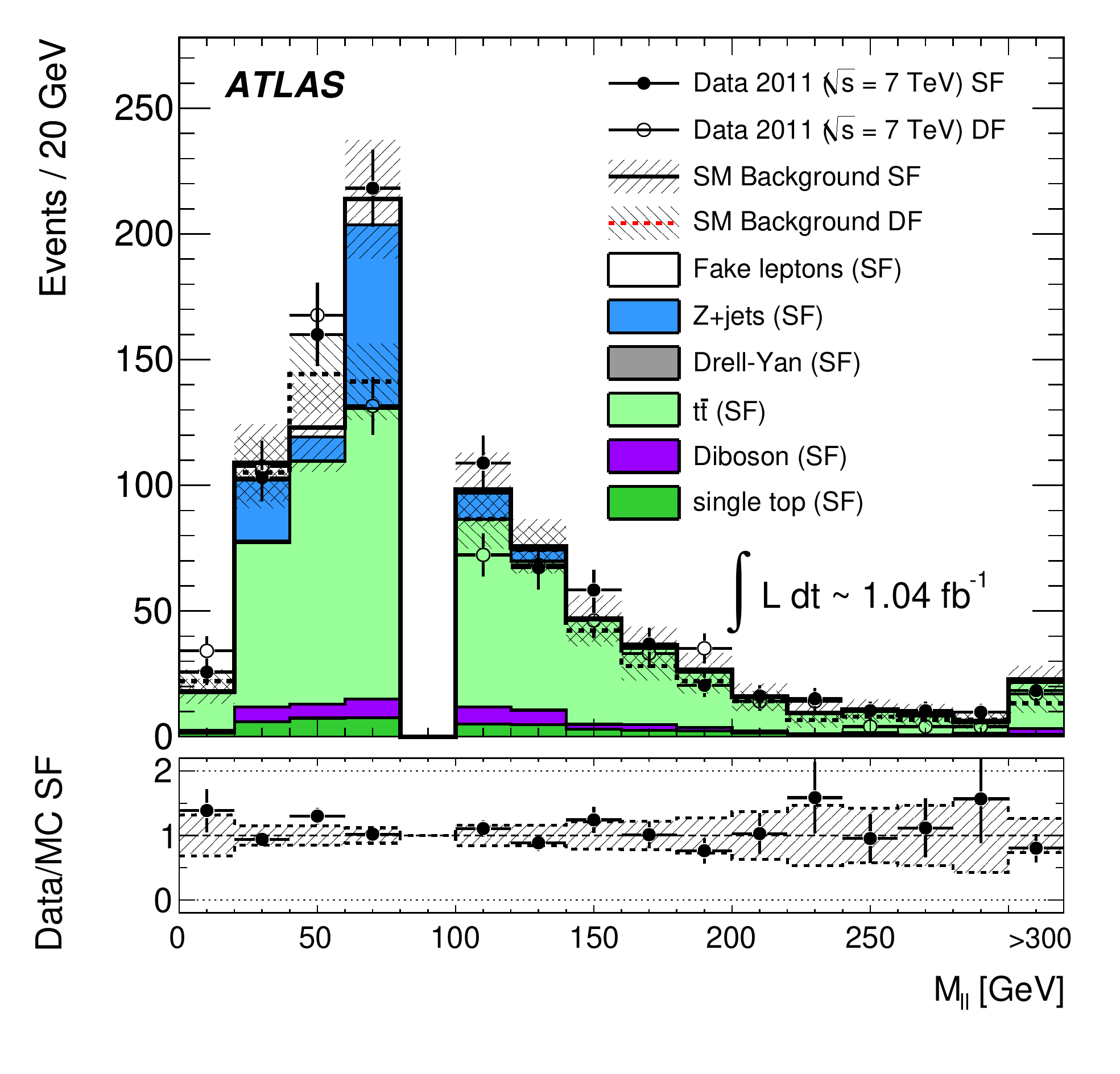}}
\subfigure{\includegraphics[scale=0.42]{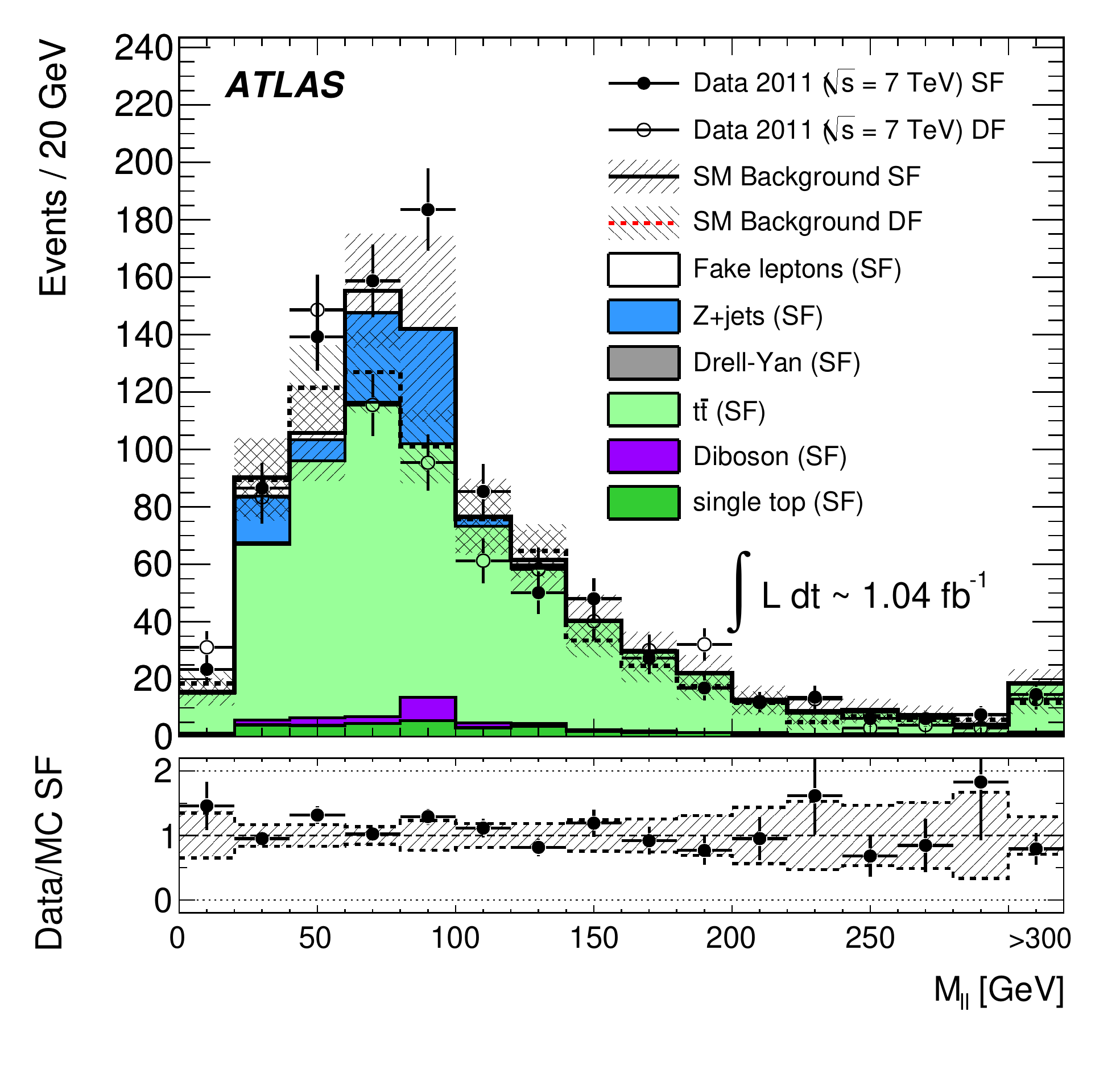}}
\caption{Distributions of the invariant mass in data together with the SM expectation for same-flavour (SF) dilepton events with $E^{\mathrm{miss}}_{\mathrm{T}}>80$~GeV after a Z-veto requirement (FS-no Z)~(a) and 2-jet requirement (FS-2j)~(b). Also shown are the different-flavour (DF) distributions. Errors on data points are statistical, while the error bands on the SM predictions represent the total uncertainties. \label{weightedMll_distributions}
}
\end{center}
\end{figure*}

The numbers of events in each signal region give $N(e^\pm e^\mp)$, $N(e^\pm \mu^\mp)$ and $N(\mu^\pm \mu^\mp)$ for each region. The invariant mass distributions of the dilepton events with high-$E^{\mathrm{miss}}_{\mathrm{T}}$ are illustrated in Fig.~\ref{weightedMll_distributions}. To quantify the consistency between the observed $\mathcal{S}$ value and the SM prediction the expected distribution of $\mathcal{S}_{b}$ in the absence of new phenomena is determined. This distribution possesses a mean given by $\bar{\mathcal{S}_{b}}$ and a width dominated by statistical fluctuations of the numbers of events observed in each channel. The distributions for ${\mathcal{S}_{b}}$ can be determined by generating pseudo-experiments. For each pseudo-experiment the mean numbers of background events in each channel and from each source are sampled, taking appropriate account of correlations between the uncertainties in the values of these means. The resulting number of background events in each channel is then used to construct a Poisson distribution from which the observed number of events in that channel is drawn. The sampled event counts in each channel are then used with Eq.~\ref{eq:FlavSubtraction} to determine a value of $\mathcal{S}_b$, taking care also to sample values of $\tau_e$, $\tau_\mu$ and $\beta$ according to their means and uncertainties. The distribution of $\mathcal{S}_{b}$ obtained from these hypothetical signal-free experiments are characterised by a mean and an RMS, as detailed in Table~\ref{Tab:Interpretation_FS_SMOnly}. The non-zero $\bar{\mathcal{S}}_{b}$ is due to the irreducible background from $Z/\gamma^*+$jets and diboson events. The assumption that the trigger selection for different flavour dilepton events is equivalent to a logical OR between the electron and muon triggers leads to a slight underestimate of the effective excess of same-flavour events in each region (greatest at 3.5$\%$ of $\mathcal{S}_{obs}$ in FS-inc, negligible in comparison to the RMS which drives the limit).

\begin{table}[!htbp]
\begin{center}
\begin{tabular}{c|c|c|c}
\hline
& & & \\[-0.35cm]
& $\mathcal{S}_{obs}$
& $\bar{\mathcal{S}}_{b}$ & RMS\\ \hline
\hline
FS-no Z &
131.6$\pm$2.5(sys) & 118.7$\pm$27.0
& 48.6 \\
FS-2j &
142.2$\pm$1.0(sys) & 67.1$\pm$28.6
& 49.0 \\
FS-inc &
-3.06$\pm$0.04(sys) & 0.7$\pm$1.6 &
4.5 \\
\hline
\end{tabular}
\end{center}
\caption{
\label{Tab:Interpretation_FS_SMOnly}The observed values of $\mathcal{S}$ in the data ($\mathcal{S}_{obs}$), and the mean and RMS of the distributions of the expected
      $\mathcal{S}_{b}$ from one million hypothetical signal-free
      pseudo-experiments.}
\end{table}

The distribution of $\mathcal{S}_b$ values obtained in this way can be used to evaluate the probability of observing a value of $\mathcal{S}$ at least as large as $\mathcal{S}_{obs}$. The width of the distribution is dominated by Poisson fluctuations in the number of events. The consistency between data and the SM expectation in each signal region is summarised in Table~\ref{Tab:Interpretation_FS}. The agreement is better than 2$\sigma$ in all cases. \begin{table}[!htbp]
\begin{center}
\begin{tabular}{c|c|c}
\hline
& & \\[-0.35cm]
& $\mathcal{S}>\mathcal{S}_{obs}$ ($\%$) & Limit
$\bar{\mathcal{S}}_{s}$ (95$\%$ CL)\\ \hline
\hline
FS-no Z & 39 & 94
\\
FS-2j & 6 & 158
\\
FS-inc & 79 &
4.5 \\
\hline
\end{tabular}
\end{center}
\caption{
\label{Tab:Interpretation_FS}Consistency of the observation with the
      SM expectation (middle column), computed as the percentage of signal-free pseudo-experiments 
      giving values of $\mathcal{S}$ greater than the observation, $\mathcal{S}_{obs}$. Observed limit (right column) on the numbers of same-flavour events from new phenomena multiplied by detector acceptances and efficiencies in each signal region. }
\end{table}

Limits are also set on $\bar{\mathcal{S}}_s$, the mean contribution to $\mathcal{S}$ from new phenomena. The statistical procedure employed follows that used to determine the consistency of the observed value of $\mathcal{S}$ with the background expectation. The pseudo-experiments are modified by adding signal event contributions to the input mean numbers of background events in each channel. An assumption must be made regarding the relative branching ratio of new processes into same-flavour and different flavour 
final states, as adding flavour uncorrelated contributions to the same-flavour and different-flavour channels increases the width of the $\mathcal{S}$ distribution. Given such an assumption, a limit can be set on $\bar{\mathcal{S}}_s$ by comparing $\mathcal{S}_{obs}$ with the distribution of $\mathcal{S}$ values obtained from the new set of signal-plus-background pseudo-experiments. If the assumption is made that the branching fractions for $e^\pm e^\mp$ and $\mu^\pm \mu^\mp$ in new physics events are identical, and the branching fraction for $e^\pm \mu^\mp$ final states is zero, then the limits tabulated in the right most column of Table~\ref{Tab:Interpretation_FS} are obtained. The most stringent limits are set in FS-inc, which requires $E^{\mathrm{miss}}_{\mathrm{T}}>250$~GeV.

\section{Summary}

This letter 
reports results of three searches for new phenomena in final states with opposite-sign and same-sign dileptons and missing transverse momentum. These searches also include signal regions that place requirements on the number and $p_\mathrm{T}$ of energetic jets in the events. There is good agreement for all signal regions between the numbers of observed events and the SM predictions. Model-independent limits are quoted on the cross section multiplied by acceptances and efficiencies for the inclusive analyses, and limits on the same-flavour excess multiplied by acceptances and efficiencies for the flavour-subtraction analysis, all of which improve on results obtained with the 2010 dataset. Cross sections in excess of 9.5~fb for opposite-sign events with missing transverse momentum greater than 250~GeV are excluded at 95$\%$ CL. In events with missing transverse energy greater than 250 GeV a limit is set on the number of same-flavour lepton pairs from new physics, multiplied by detector acceptance and efficiency, of 4.5. Cross sections in excess of 10.2~fb for same-sign events, with missing transverse momentum greater than 100~GeV, are excluded at 95$\%$ CL. Additionally, new limits have been presented on the chargino mass in direct electroweak gaugino production modes using simplified models. Charginos with masses up to 200~GeV are excluded, under the assumptions of these models.

\section{Acknowledgements}

We wish to thank CERN for the efficient commissioning and operation of the LHC during this initial high-energy data-taking period as well as the support staff from our institutions without whom ATLAS could not be operated efficiently.

We acknowledge the support of ANPCyT, Argentina; YerPhI, Armenia; ARC, Australia; BMWF, Austria; ANAS, Azerbaijan; SSTC, Belarus; CNPq and FAPESP, Brazil; NSERC, NRC and CFI, Canada; CERN; CONICYT, Chile; CAS, MOST and NSFC, China; COLCIENCIAS, Colombia; MSMT CR, MPO CR and VSC CR, Czech Republic; DNRF, DNSRC and Lundbeck Foundation, Denmark; ARTEMIS, European Union; IN2P3-CNRS, CEA-DSM/IRFU, France; GNAS, Georgia; BMBF, DFG, HGF, MPG and AvH Foundation, Germany; GSRT, Greece; ISF, MINERVA, GIF, DIP and Benoziyo Center, Israel; INFN, Italy; MEXT and JSPS, Japan; CNRST, Morocco; FOM and NWO, Netherlands; RCN, Norway;  MNiSW, Poland; GRICES and FCT, Portugal;  MERYS (MECTS), Romania;  MES of Russia and ROSATOM, Russian Federation; JINR; MSTD, Serbia; MSSR, Slovakia; ARRS and MVZT, Slovenia; DST/NRF, South Africa; MICINN, Spain; SRC and Wallenberg Foundation, Sweden; SER,  SNSF and Cantons of Bern and Geneva, Switzerland;  NSC, Taiwan; TAEK, Turkey; STFC, the Royal Society and Leverhulme Trust, United Kingdom; DOE and NSF, United States of America.  

The crucial computing support from all WLCG partners is acknowledged gratefully, in particular from CERN and the ATLAS Tier-1 facilities at TRIUMF (Canada), NDGF (Denmark, Norway, Sweden), CC-IN2P3 (France), KIT/GridKA (Germany), INFN-CNAF (Italy), NL-T1 (Netherlands), PIC (Spain), ASGC (Taiwan), RAL (UK) and BNL (USA) and in the Tier-2 facilities worldwide.

\bibliographystyle{atlasnote}
\bibliography{paper_2lepton}

\providecommand{\href}[2]{#2}\begingroup\raggedright\begin{thebibliography}{10}

\bibitem{susy}
{Yu.A. Golfand and E.P. Likhtman, JETP Lett. 13 (1971) 323; A. Neveu and J.H.
  Schwartz, Nucl. Phys. B31 (1971) 86; A. Neveu and J.H. Schwartz, Phys. Rev.
  D4 (1971) 1109; P. Ramond, Phys. Rev. D3 (1971) 2415; D.V. Volkov and V.P.
  Akulov, Phys. Lett. B46 (1973) 109; J. Wess and B. Zumino, Phys. Lett. B49
  (1974) 52; J. Wess and B. Zumino, Nucl. Phys. B70 (1974) 39.}

\bibitem{r-parity}
{P.~Fayet, Phys. Lett. B {\bf 69} (1977) 489; G.~R.~Farrar and P.~Fayet, Phys.
  Lett. B {\bf 76} (1978) 575.}

\bibitem{Hinchliffe:1996iu}
I.~Hinchliffe, F.~E. Paige, M.~D. Shapiro, J.~Soderqvist, and W.~Yao,
\href{http://dx.doi.org/10.1103/PhysRevD.55.5520}{Phys. Rev. {\bf D55} (1997)
  5520--5540.}

\bibitem{Allanach:2000kt}
B.~C. Allanach, C.~G. Lester, M.~A. Parker, and B.~R. Webber,
JHEP {\bf 09} (2000)  004.

\bibitem{Costanzo:2009mq}
D.~Costanzo and D.~R. Tovey,
\href{http://dx.doi.org/10.1088/1126-6708/2009/04/084}{JHEP {\bf 04} (2009)
  084.}

\bibitem{atlas2lep1}
{ATLAS} Collaboration, Eur.Phys.J. {\bf C71} (2011)  1682.

\bibitem{atlas2lep2}
{ATLAS} Collaboration,
\href{http://dx.doi.org/10.1140/epjc/s10052-011-1647-9}{Eur.Phys.J. {\bf C71}
  (2011)  1647.}

\bibitem{cms2lepOS}
{CMS Collaboration, JHEP 06 (2011) 026}.

\bibitem{cms2lepSS}
{CMS Collaboration, JHEP 06 (2011) 077}.

\bibitem{Aad:2008zzm}
{ATLAS Collaboration, JINST {\bf 3} (2008) S08003}.

\bibitem{lumicorrect}
ATLAS Collaboration, Eur. Phys. J. C \textbf{71} (2011) 1630 and
  ATLAS-CONF-2011-011-116 {\url{http://cdsweb.cern.ch/record/1376384} (2011)}.

\bibitem{mcatnlolong}
{S.~Frixione, B.~R.~Webber, JHEP 06 (2002) 029; S.~Frixione, P.~Nason and
  B.~R.~Webber, JHEP 08 (2003) 007; S.~Frixione, E.~Laenen and P.~Motylinski,
  JHEP 03 (2006) 092}.

\bibitem{cteq66}
{P.~M.~Nadolsky et al., Phys.~Rev.~D78 (2008) 013004}.

\bibitem{Mangano:2002ea}
M.~L. Mangano, M.~Moretti, F.~Piccinini, R.~Pittau, and A.~D. Polosa,
JHEP {\bf 07} (2003)  001.

\bibitem{herwig}
G.~Corcella et al.,
JHEP {\bf 01} (2001)  010.

\bibitem{madgraph}
{J. Alwall et al., JHEP {\bf 06} (2011) 128}.

\bibitem{Butterworth:1996zw}
J.~Butterworth, J.~Forshaw, and M.~Seymour,
Z. Phys. {\bf C72} (1996)  637--646.

\bibitem{powheg}
S.~Frixione, P.~Nason, and C.~Oleari, JHEP {\bf 11} (2007)  070.

\bibitem{Kersevan:2004yg}
B.~P. Kersevan and E.~Richter-Was, {arXiv:0405247} (2004).

\bibitem{herwigpp}
{M.~Bahr et al., Eur. Phys. J. C58 (2008) 639}.

\bibitem{prospino}
{W. Beenakker, M. Klasen, M. KrŠmer, T. Plehn, M. Spira, and P. M. Zerwas,
Phys. Rev. Lett. 83, 3780Ð3783 (1999).}

\bibitem{Sherstnev:2007nd}
A.~Sherstnev and R.~Thorne,
\href{http://dx.doi.org/10.1140/epjc/s10052-008-0610-x}{Eur. Phys. J. {\bf C55}
  (2008)  553--575}.

\bibitem{mc10a}
{ATLAS Collaboration, ATL-PHYS-PUB-2010-014
  \url{https://cdsweb.cern.ch/record/1303025} (2010) and ATLAS-CONF-2010-031
  \url{http://cdsweb.cern.ch/record/1277665} (2010)}.

\bibitem{geant4}
{S.~Agostinelli~et~al.,~Nucl.~Instr.~Meth.~A {\bf 506}~(2003)~250}.

\bibitem{atlassimulation}
{ATLAS Collaboration, G. Aad et al., Eur. Phys. J. C {\bf 70} (2010) 823}.

\bibitem{Aad:2010yt}
{{ATLAS}} Collaboration, \href{http://dx.doi.org/10.1007/JHEP12(2010)060}{JHEP
  {\bf 1012} (2010)  060}.

\bibitem{antikt}
{M. Cacciari, G.P. Salam and G. Soyez, JHEP {\bf 04} (2008) 063; M. Cacciari
  and G. P. Salam, Phys. Lett. B {\bf 641} (2006) 57}.

\bibitem{atlas-jet-clean}
{ATLAS Collaboration, ATLAS-CONF-2010-038
  \url{http://cdsweb.cern.ch/record/1277678} (2010)}.

\bibitem{Tovey:2008}
D.~Tovey, \href{http://dx.doi.org/10.1088/1126-6708/2008/04/034}{JHEP {\bf
  0804} (2008)  034}.

\bibitem{Polesello:2009rn}
G.~Polesello and D.~Tovey,
  \href{http://dx.doi.org/10.1007/JHEP03(2010)030}{JHEP {\bf 1003} (2010)
  030}.

\bibitem{atlas-topxsec}
{ATLAS} Collaboration,
\href{http://dx.doi.org/10.1140/epjc/s10052-011-1577-6}{Eur. Phys. J. {\bf C71}
  (2011)  1577}.

\bibitem{jes}
{ATLAS Collaboration, ATLAS-CONF-2011-032
  \url{http://cdsweb.cern.ch/record/1337782} (2011)}.

\bibitem{Read:2002hq}
A.~L. Read, \href{http://dx.doi.org/10.1088/0954-3899/28/10/313}{J.Phys.G {\bf
  G28} (2002)  2693--2704}.

\bibitem{simplified}
{J. Alwall, P. Schuster and N. Toro, Phys Rev D79 (2009) 075020; D. Alves et
  al., arXiv:1105.2838 (2011)}.
  
\bibitem{Abazov:2009zi}
{{ D0} Collaboration, { Phys. Lett.} \textbf{ B680} (2009) 34.}

\bibitem{Forrest:2009gm}
{{ CDF} Collaboration,
  \href{http://www.arXiv.org/abs/0910.1931}{\texttt{ arXiv:0910.1931}}.

\bibitem{lepsusy}
\href
  {http://lepsusy.web.cern.ch/lepsusy/www/sleptons_summer04/slep_final.html} {{
  LEP2 SUSY Working Group} Collaboration}}

\bibitem{Heister:2001nk}
 {{ ALEPH} Collaboration, { Phys.
  Lett.} \textbf{ B526} (2002) 206.}

\bibitem{Heister:2003zk}
 {{ ALEPH} Collaboration,{ Phys. Lett.} \textbf{ B583} (2004) 247.}
  
\bibitem{Abdallah:2003xe}
{{ DELPHI} Collaboration, { Eur. Phys. J.} \textbf{ C31} (2003)}

\bibitem{Achard:2003ge}
{{ L3} Collaboration,{ Phys. Lett.} \textbf{ B580} (2004) 37.}

\bibitem{Abbiendi:2003ji}
{{ OPAL} Collaboration, { Eur. Phys. J.} \textbf{ C32} (2004)}


\end{thebibliography}\endgroup

\clearpage
\onecolumn

\begin{flushleft}
{\Large The ATLAS Collaboration}

\bigskip

G.~Aad$^{\rm 48}$,
B.~Abbott$^{\rm 111}$,
J.~Abdallah$^{\rm 11}$,
A.A.~Abdelalim$^{\rm 49}$,
A.~Abdesselam$^{\rm 118}$,
O.~Abdinov$^{\rm 10}$,
B.~Abi$^{\rm 112}$,
M.~Abolins$^{\rm 88}$,
H.~Abramowicz$^{\rm 153}$,
H.~Abreu$^{\rm 115}$
% ...
\bigskip

% ...
$^{10}$ Institute of Physics, Azerbaijan Academy of Sciences, Baku, Azerbaijan\\
$^{11}$ Institut de F\'isica d'Altes Energies and Departament de F\'isica de la Universitat Aut\`onoma  de Barcelona and ICREA, Barcelona, Spain\\
% ...
$^{48}$ Fakult\"{a}t f\"{u}r Mathematik und Physik, Albert-Ludwigs-Universit\"{a}t, Freiburg i.Br., Germany\\
$^{49}$ Section de Physique, Universit\'e de Gen\`eve, Geneva, Switzerland\\
% ...
$^{88}$ Department of Physics and Astronomy, Michigan State University, East Lansing MI, United States of America\\
% ...
$^{111}$ Homer L. Dodge Department of Physics and Astronomy, University of Oklahoma, Norman OK, United States of America\\
$^{112}$ Department of Physics, Oklahoma State University, Stillwater OK, United States of America\\
% ...
$^{115}$ LAL, Univ. Paris-Sud and CNRS/IN2P3, Orsay, France\\
% ...
$^{118}$ Department of Physics, Oxford University, Oxford, United Kingdom\\
% ...
$^{153}$ Raymond and Beverly Sackler School of Physics and Astronomy, Tel Aviv University, Tel Aviv, Israel\\
% ...
\end{flushleft}

\end{document}